\documentclass[aip,amsmath,amssymb, reprint]{revtex4-1}
\usepackage{physics}
\usepackage{amsmath}
\usepackage{mathptmx,bm}
\usepackage{graphicx}
\usepackage{dcolumn}
\usepackage{tabularx}
\usepackage{epsf}
\usepackage{color}
\usepackage{hyperref}
\hypersetup{breaklinks,colorlinks,linkcolor=blue,citecolor=blue,urlcolor=blue}

\usepackage{epstopdf}%
\usepackage{verbatim}
\usepackage[english]{babel}
\newcommand{\bea}{\begin{eqnarray}}
	\newcommand{\eea}{\end{eqnarray}}
\usepackage{amsfonts}
\usepackage{graphicx}
\usepackage{dcolumn}
\usepackage{bm}
\usepackage[utf8]{inputenc}
\usepackage[T1]{fontenc}
\usepackage{mathptmx}
\usepackage{etoolbox}
\usepackage{multirow}

\makeatletter
\def\@email#1#2{%
	\endgroup
	\patchcmd{\titleblock@produce}
	{\frontmatter@RRAPformat}
	{\frontmatter@RRAPformat{\produce@RRAP{*#1\href{mailto:#2}{#2}}}\frontmatter@RRAPformat}
	{}{}
}%
\makeatother

\begin{document}

\title{Challenges in molecular dynamics simulations of heat exchange statistics}



\author{Jonathan J. Wang}
\affiliation{Chemical Physics Theory Group, Department of Chemistry, University of Toronto, 80 Saint George St., Toronto, Ontario M5S 3H6, Canada}

\author{Matthew Gerry}
\affiliation{Department of Physics, University of Toronto, 60 Saint George St., Toronto, Ontario M5S 1A7, Canada}

\author{Dvira Segal}
\affiliation{Chemical Physics Theory Group, Department of Chemistry, University of Toronto, 80 Saint George St., Toronto, Ontario M5S 3H6, Canada}

\affiliation{Department of Physics, University of Toronto, 60 Saint George St., Toronto, Ontario M5S 1A7, Canada}
\email{dvira.segal@utoronto.ca}

\date{\today}

\begin{abstract}
We study heat exchange in temperature-biased metal-molecule-metal molecular junctions
by employing the LAMMPS atomic molecular dynamics simulator.
Generating the nonequilibrium steady state with Langevin thermostats at the boundaries of the junction,
we show that the {\it average} heat current across a gold-alkanedithiol-gold
nanojunction behaves correctly-physically, 
with the thermal conductance value matching the literature.
In contrast, the {\it full probability distribution function} for heat  exchange, as generated by the simulator, violates the fundamental fluctuation symmetry for entropy production.
We trace this failure back to the implementation of the thermostats and the expression used to calculate the heat exchange.
To rectify this issue and produce the correct statistics, we introduce single-atom thermostats as an alternative to conventional many-atom thermostats. 
Once averaging heat exchange over the hot and cold thermostats, this approach successfully generates the correct probability distribution function, which we use to study the behavior of both the average heat current and its noise.
We further examine the thermodynamic uncertainty relation in 
the molecular junction and show that it holds, albeit demonstrating nontrivial trends. 
Our study points to the need to carefully implement nonequilibrium molecular dynamics solvers in 
atomistic simulation software tools for future investigations of noise phenomena in thermal transport. 
\end{abstract}

\maketitle
\section{Introduction}

Nanoscale thermal transport has been increasingly studied in relation to different applications, including electronics, thermoelectrics, and plasmonics
\cite{Pop10,Baowen12,Luo13,Rev14,Leitner15,RevA,Yoon20,BaowenR21,Baowen22,RevG}. 
Considering molecular-based technologies, recent experiments have focused on characterizing and elucidating thermal transport mechanisms in self-assembled monolayers (SAMs)\cite{Wang06,Dlott07,Cahill12,GotsmannExp14,Shub15,Shub17} and single-molecule junctions\cite{CuiExp19,GotsmannExp19,GotsmannExp23}. 
Focusing on the latter, it is important to recognize that the instrumentation involved is  highly challenging due to the minuscule currents at play.
For instance, heat currents measured in single alkane junctions are in the picowatts range; such measurements are done by reading electronic probes acting at the interface, rather than within the molecule itself \cite{CuiExp19}.

Studies of thermal transport in single molecules provide fundamental understanding on transport mechanisms as well as guidelines for advancing energy technologies. 
However, due the experimental challenge involved in such studies,
simulations play a central role in supporting and directing experiments.
In this context, classical molecular dynamics (MD) simulations stand out as a practical and reasonably accurate tool for characterizing thermal transport behavior, especially at elevated temperatures. 
Although these simulations are obviously approximate in modelling interactions between atoms, compared to 
first-principle based approaches, the use of empirical potentials significantly reduces computational demands
and allow probing questions such as how the introduction of many body effects allow the transition from anomalous to normal conduction \cite{Dhar,Lepri}. 

Given their structural simplicity, thermal transport in
gold-alkanedithiol-gold single-molecule junctions has been examined in detail both experimentally \cite{CuiExp19,GotsmannExp19,GotsmannExp23} and computationally \cite{Dvira2003,Pawel11,Pauly16,Pauly18,Roya19,Nitzan20, NitzanD20,Lu2021,Nitzan22,Pauly23,JW-HeatMD}.  
Notably, simulations of thermal conductance of these systems showed that transport was ballistic (non-dissipative) thus approximately length-independent, in good agreement with experimental results.

\begin{figure}
    \centering
 \includegraphics[width=\columnwidth]{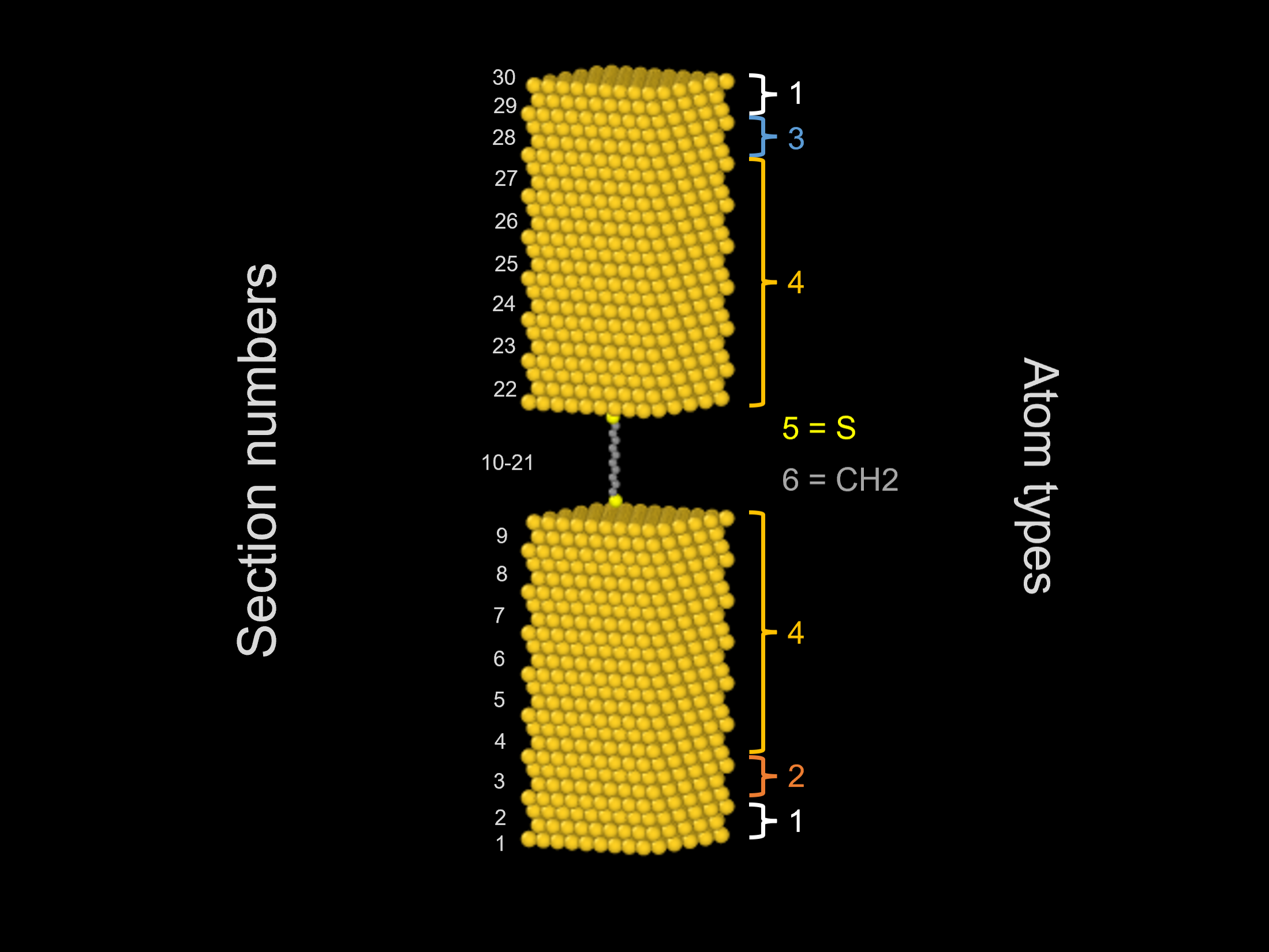}
\caption{{\bf Visualization of the system simulated in this work} (OVITO program\cite{OVITO}). 
The setup includes two Au metals with 2160 atoms each, connected by an alkanedithiol molecule with 10 carbon atoms. In this setup, Langevin baths are enacted close to the ends of Au leads, away from the conducting molecule.
Atom types appearing at the right side distinguishes between regions of Au leads as well as the molecule as defined in LAMMPS. 
Type 1 atoms are fixed Au atoms that hold the system in place ($270$ atoms  at each side). 
Type 2 and 3 atoms are Au regions that are thermostated with hot and cold baths, respectively ($270$ atoms  at each side). 
Type 4 atoms encompass the remaining moving Au atoms. 
Type 5 and 6 atoms comprise the single alkane molecule, distinguishing between the S and the unified CH$_2$ atoms.
Section numbers (left) denote regions over which temperature is averaged to produce the simulation temperature profile. 
}
\label{Fig1}
\end{figure}

Nonequilibrium atomistic MD simulations of heat transport have been traditionally focused on the analysis of the average heat current passing through the junction.
In contrast, investigations of {\it full counting statistics} have been confined to simplistic one-dimensional model systems of beads and springs \cite{Fluc1,Fluc2,Fluc3,Fluc4,UdoRev12,Petrosyan13,Lepri98, Dhar08, Naim}. 
Identifying this gap, our objective in this paper is to explore the full probably distribution function of heat exchange, $P(Q_{\tau})$ within a 
detailed, experimentally relevant atomistic model of a thermal transport junction, as illustrated in Fig. \ref{Fig1}.
Here, $Q_{\tau}$ is defined as the net heat exchange between the hot reservoir and the system within a time $\tau$; 
one can also calculate the corresponding heat exchange at the cold end.  $\tau$ stands for a time interval long enough to eliminate dependence on initial conditions. 


Our motivation for studying the properties of $P(Q_{\tau})$ is twofold.
Primarily, this would allow testing and improving the simulation algorithm through the verification of the heat exchange fluctuation theorem, which holds in nonequilibrium steady state \cite{Fluc1,Fluc2,Fluc3,Fluc4,UdoRev12,Petrosyan13,Lepri98, Dhar08, Naim},
\bea
\frac{P(Q_{\tau})}{P(-Q_{\tau})} = e^{\Delta \beta Q_{\tau}},
\label{eq:PQ}
\eea
Here, $\Delta\beta\equiv \frac{1}{k_B}\left(\frac{1}{T_\text{cold}}-\frac{1}{T_\text{hot}}\right)$ is the difference between 
the inverse temperatures of the two electrodes 
with $k_B$ as the Boltzmann's constant (a more precise discussion of $\Delta \beta$ is included in Sec. \ref{Method-sec}). 
$\Delta \beta Q_{\tau}$ is the entropy production in the reservoirs during time $\tau$.
Eq. (\ref{eq:PQ}) is a special case of the more general statement on the symmetry of entropy production in nonequilibrium steady state, leading to the second law of thermodynamics \cite{Fluc1,Fluc2,Fluc3,Fluc4,UdoRev12,Petrosyan13,Lepri98, Dhar08}.
Confirming the validity of the relation (\ref{eq:PQ}) through simulations serves to attest that the simulation method generates data which obeys physical principles.
This in turn enables the investigation of the nature of fluctuations and their influence on the operation of thermal devices, which is our second motivation for studying the behavior of this distribution function.
While such investigations were carried out on model systems such as the nonequilibrium spin-boson model \cite{Nicolin11,Nicolin11PRB,Segal14,SegalNJP,SegalPI,Cao15,Brandes16,Aurell20,He21}, simulations of heat current fluctuations in detailed atomistic nanojunctions are lacking.

In particular, cost (dissipation) - precision (noise) tradeoffs in nanoscale systems gained enormous interest in the last decade \cite{TURrev1,TURrev2,TURUdo1,TURUdo2,TURGing,TURLang,DechantSasa,TURheat,TURhE,TURrun}.
The thermodynamic uncertainty relation (TUR) was proved for continuous-time Markov state processes \cite{TURGing}, and subsequently for overdamped Langevin dynamics \cite{TURLang,DechantSasa}. However, the original TUR does not universally hold under underdamped Langevin dynamics \cite{TURrun}. 
The study of nonequilibrium noise in molecular junctions and the analysis of the TUR beyond overdamped dynamics hold significance both from a fundamental perspective and for applications.

In this work, we  study heat exchange within a gold-alkanedithiol-gold single-molecule junction using classical MD simulations. We begin by employing the standard procedure of nonequilibrium molecular dynmaics (NEMD) simulations as implemented in the LAMMPS software.
Our work reveals that the conventional approach of utilizing many-atom Langevin thermostats at the boundaries of the junction
results in a distribution function $P(Q_{\tau})$ that fails to satisfy the fluctuation theorem for heat exchange.
While the NEMD method correctly provides the average heat current, it does not capture the correct statistical behavior of heat fluctuations. 

Several algorithmic issues contribute to this discrepancy, including the adoption of a problematic definition for heat exchange \cite{Naim} and the use of deterministic solvers for differential equations, instead of stochastic solvers. Notably, our analysis identifies the application of many-atom thermostats as another significant contributor to the breakdown of the fluctuation symmetry. As such, in the second part of the paper we introduce {\it single-atom thermostats}, which thermalize a single atom at each boundary. This setup
rectifies the statistics,  leading to distribution functions $P(Q_{\tau})$ that satisfy the fluctuation symmetry.
Importantly, we confirm that our modified thermostat approach delivers the correct thermal conductance coefficient and produces physical temperature profiles.
We also study the current noise and observe nontrivial and nonmonotonic behavior in the TUR ratio. 
These findings will guide measurements of full counting statistics of heat exchange in single-molecule thermal junctions.

%

The paper is organized as follows.
In Sec. \ref{Method-sec} we outline the NEMD simulation methodology and explain the analysis required to build the distribution function $P(Q_{\tau})$. 
We present results from simulations in two sections: In Sec. \ref{many-sec}, we discuss calculations of the full probability distribution function for heat exchange when using conventional many-atom thermostats, demonstrating violations to the fluctuation symmetry. 
This deficiency is addressed and resolved in Sec. \ref{single-sec}, by implementing single-atom thermostats. 
We summarize our work in Sec. \ref{Sum-sec} and discuss potential future directions.


\section{Setup and Simulation Technique}
\label{Method-sec}

\subsection{Setup and procedure}
We study heat transfer through a single molecule positioned between gold leads, as illustrated in Fig. \ref{Fig1}.
In our work, we ignore the direct electronic contribution to heat flow, and focus on the phononic-vibrational contribution to it.
Since we take into account interatomic interactions beyond the harmonic force field, we rely on classical MD as our tool of choice.

Our primary objectives are to produce MD data 
for molecular junctions consistent with the heat exchange fluctuation symmetry, and use it to study the heat current and its fluctuations.
As such, the specific molecule chosen is not critical to our investigation.
Given that notable 
experimental studies have focused on thermal transport within alkane chains \cite{CuiExp19,GotsmannExp19}, we have opted in our study to employ an alkanedithiol molecule with 10 carbon atoms.

We carry out simulations using the Large-scale Atomic/Molecular Massively Parallel Simulator \cite{LAMMPS} (LAMMPS), and we visualize the system through OVITO \cite{OVITO}. 
To study phononic-vibrational heat transport, we employ the NEMD 
simulation approach. In this method, a temperature bias is imposed through the boundaries, and it induces a net heat current across the junction.
The gold-alkanedithiol-gold system setup and the empirical interatomic potentials utilized were described in detail in Ref. \citenum{JW-HeatMD}. 
Briefly, before the production run, the system undergoes equilibration in the NPT then NVT ensembles to  
attain zero pressure and to reach a target temperature $\bar{T}$. 
During the NVE simulation production run,
Langevin thermostats are applied to opposite segments of the gold, one set at the high temperature $T_\text{hot}$, the other at the low temperature $T_\text{cold}$, with the imposed bias $\Delta T = T_\text{hot}-T_\text{cold}$. 

Specifically, the gold atoms that are coupled to the baths follow the Langevin equation of motion,
which ensures controlled temperature conditions in that boundary,  
$m_\text{Au}\frac{d\vec v_n}{dt} = \vec F_n(t) - \gamma m_\text{Au} \vec v_n + \vec \xi_n(t)$.
Here, $\vec F_n$ is the deterministic force acting on atom $n$, derived from the interatomic potentials at time $t$.
$\vec v_n$ is the velocity vector of atom $n$ with mass $m_\text{Au}$.
As for the Langevin terms,
$\gamma$ is the friction, or damping coefficient (dimension of inverse time) and $\vec \xi_n(t)$ is a  Gaussian stochastic force applied to the $n$th atom with a delta-time  correlation function, $\langle \xi_{n,j}(t)\xi_{n,k}(t')\rangle = 2\gamma k_BTm_\text{Au}\delta(t-t')\delta_{k,j}$; $T$ is the temperature of the thermostat and $k,j$ refer to the different spatial coordinates for each atom $n$.  
Input parameters to this simulation are 
the temperatures $T_\text{hot}$ and $T_\text{cold}$ and the damping rate $\gamma$. 
In our simulations, we employed a timestep $\Delta t = 1$ fs and performed production
simulations over a total simulation time ranging from 10 ns (for studying thermal conductance) to 50 ns (for conducting fluctuation analysis).

The instantaneous heat current at the hot side is given by 
\bea
&&J_\text{hot}(t)  = 
\nonumber\\
&&-\gamma \sum_n\sum_{j=x,y,z} \frac{[p_{n,j}(t)]^2}{m_\text{Au} } + \frac{1}{m_\text{Au}}\sum_n\sum_{j=x,y,z} \xi_{n,j}(t) p_{n,j}(t),
\nonumber\\
\label{eq:Jdef}
\eea
%
where the sum is performed over all degrees of freedom that are being thermostated, with $n$ an atom index and $j=x,y,z$ the direction. Here, $p_{n,j}(t)$ is the momentum of a particle and $\xi_{n,j}(t)$ the applied random force. 
A similar expression can be written to calculate the instantaneous heat current at the cold end, $J_\text{cold}$.
The cumulative heat exchange is obtained by summing up the heat current over a certain interval. For example, the amount of heat exchange with the hot Langevin bath is
\bea
Q^\text{hot}_{\tau}= \int_{\tau_0}^{\tau_0+\tau} J_\text{hot}(t) dt.
\label{eq:Qhot}
\eea
Here, $\tau_0$ is selected after the nonequilibrium steady state sets in. 
Our sign convention is such that heat input from the hot Langevin bath to the thermostated atoms is defined as positive. 
As for the cold reservoir, we use the opposite sign convention. 

\subsection{Data analysis}

\begin{figure}[htbp] 
\centering
\includegraphics[width=\columnwidth]{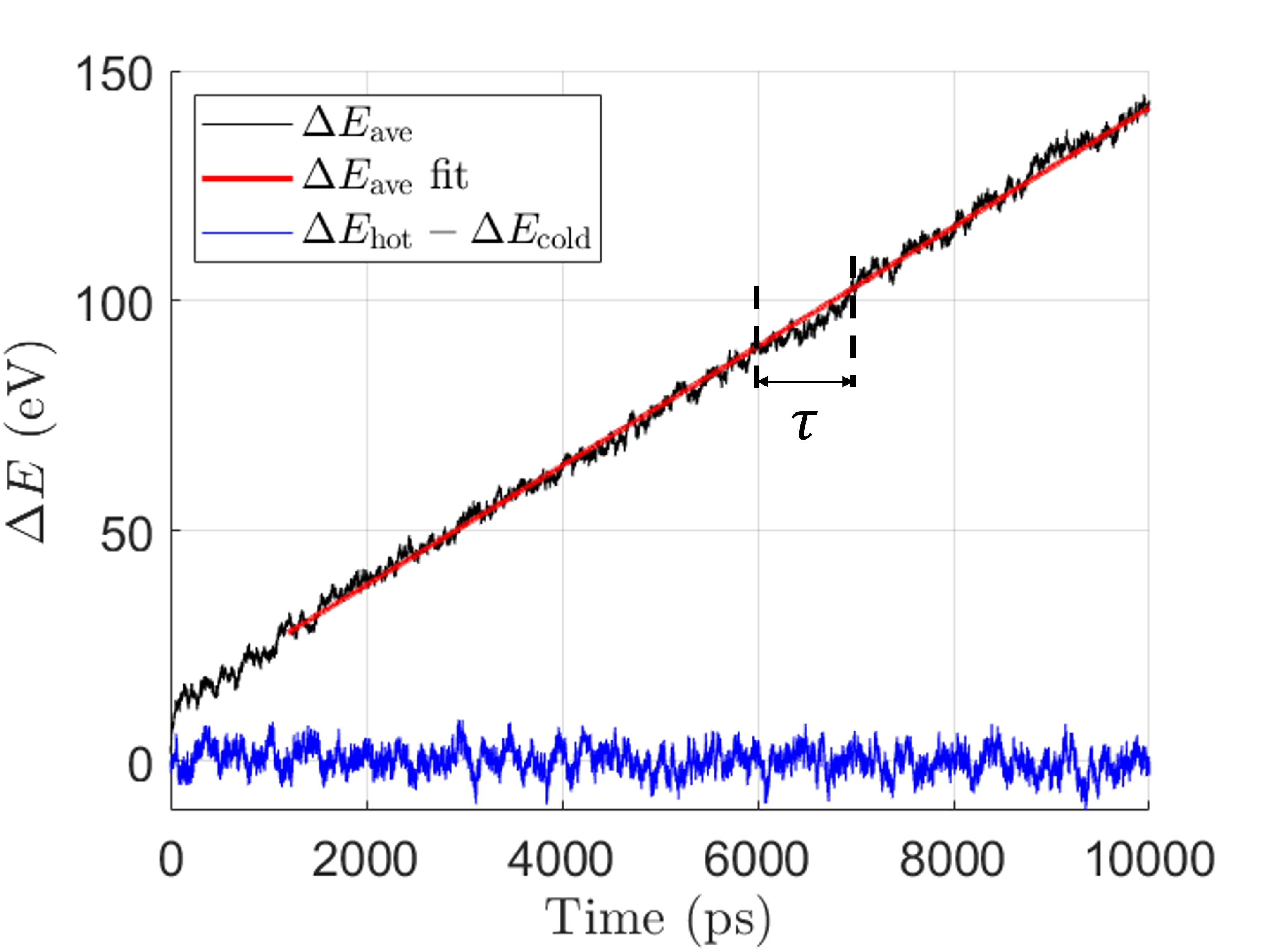}
\caption{{\bf Cumulative energy exchange of the system with the Langevin thermostats presented as a function of time.}
$\Delta E_\text{ave}(t)$ is the average of $\Delta E_\text{hot}(t)$ and $\Delta E_\text{cold}(t)$;
the full red line represents a linear fit from which the current $J$ is extracted. 
An example of the definition of integration time interval $\tau$ is shown. 
In this plot, $\tau$ is approximately 1000 ps, and the difference between $\Delta E$ values at the dashed lines give $Q_{\tau}$. 
We additionally display energy conservation in our simulated system with $\Delta E_\text{hot}(t)-\Delta E_\text{cold}(t)$. Simulation parameters are $\bar{T} = 300$ K, $\Delta T = 50$ K, $\gamma^{-1} = 0.04$ ps. }
\label{Fig2}
\end{figure}


\begin{figure*}[htbp] 
 \centering
 \includegraphics[width=2\columnwidth]{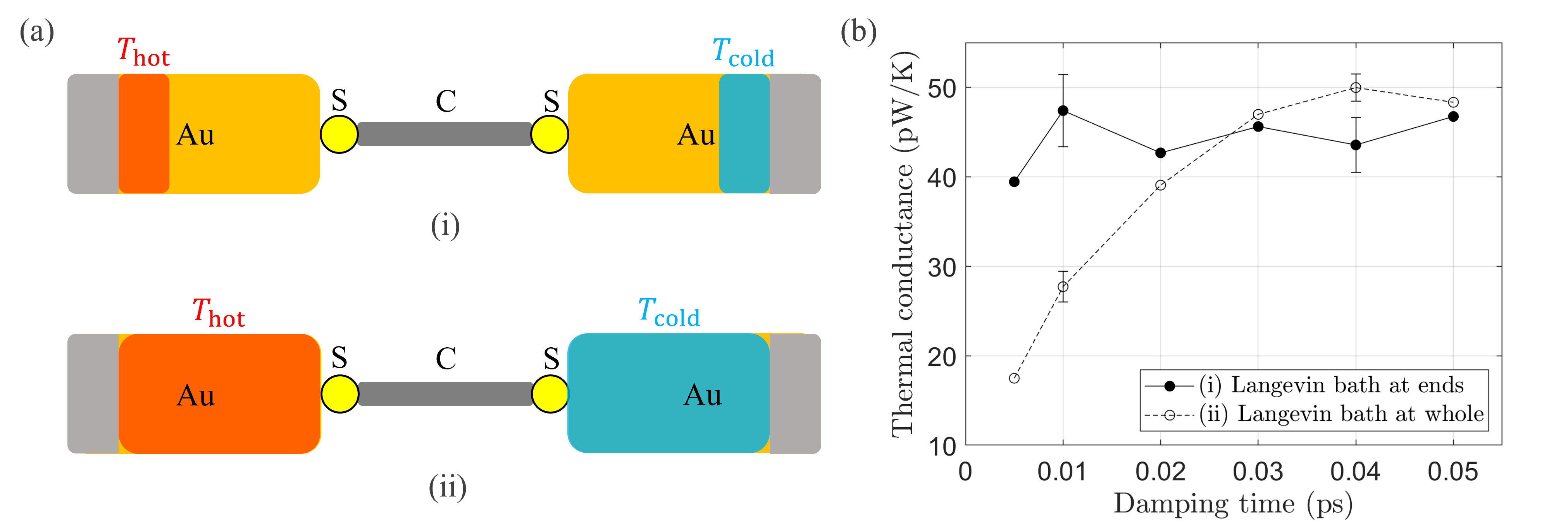}
\caption{
{\bf Nonequilibrium molecular dynamics simulations with different many-atom Langevin thermostats at the two ends.}
{\bf (a)} Each (cold, hot) thermostat either apply to (i) only a certain region of the metal further away from the molecule 
or to (ii) all moving Au atoms.
The light gray blocks at the ends indicate fixed Au that hold the system in place. 
{\bf (b)} The resulting thermal conductance as a function of Langevin damping time parameter, the  
inverse of $\gamma$ in the Langevin equation. The temperatures are set with $\bar{T} = 300$ K, $\Delta T = 50$ K.
For $\gamma^{-1}=0.01$ and 0.04 ps we performed each calculation three times and the error bars reflect the standard deviation of results. 
}
\label{Fig3}
\end{figure*}

The output data from LAMMPS provides us with the {\it cumulative} energy absorbed by the system from the hot and cold Langevin thermostats, separately.
This data is collected as a function of time \cite{LAMMPS}.  
We denote the total energy exchange with the hot and cold thermostats up to time $t$ by
$\Delta E_\text{hot}(t)$ and $\Delta E_\text{cold}(t)$, respectively, where, e.g.,
$\Delta E_\text{hot}(t)= \int_{0}^{t} J_\text{hot}(t') dt'$. 
%
In the steady state limit, the time-averaged energy emitted from the hot bath should be equal to 
the energy absorbed by the cold bath. Indeed, as seen in Fig. \ref{Fig2}, 
the difference $\Delta E_\text{hot}(t) - \Delta E_\text{cold}(t)$ hovers around zero, in line with expectations (recall our sign convention, given below Eq. (\ref{eq:Qhot})).
Furthermore, in steady state the averaged heat current across the junction 
can be calculated equivalently at either contact as
$J_\text{hot} = \frac{\Delta E_\text{hot}(t)}{t}$,
$J_\text{cold} = \frac{\Delta E_\text{cold}(t)}{t}$,
as well as from their average 
%
$J_\text{ave} = \frac{\Delta E_\text{ave}(t)}{t}$,
where $\Delta E_\text{ave}(t) = \frac{\Delta E_\text{hot}(t)+\Delta E_\text{cold}(t)}{2}$.
%
In Fig. \ref{Fig2}, we display the cumulative energy exchange $\Delta E_\text{ave}(t)$ as a function of time.
As expected, it grows linearly in time, and
we extract the heat current as the slope of this curve. 


The thermal conductance of the junction, denoted by $G$, is defined  from the relationship $J = G\Delta T_\text{Au-Au}$. Here $J$ stands for the current, which can be evaluated at either contacts, or computed from the average, with $\Delta T_\text{Au-Au}$ representing the actual temperature difference between the metals. 
We present such simulations in Fig. \ref{Fig3} (discussed in details in Sec. \ref{many-sec}). 
In Ref. \citenum{JW-HeatMD}, we discussed both the molecular conductance and the junction's conductance.

The role of the thermostats is to drive the two different metals towards distinct temperatures,
 $T_\text{hot}$ and $T_\text{cold}$. We define $\Delta T=T_\text{hot}-T_\text{cold}$ as the {\it imposed} temperature difference. 
 We further introduce two other measures for our analysis.
 As mentioned above, $\Delta T_\text{Au-Au}$ stands for the temperature difference as observed on gold atoms located at the gold layer closest to the molecule.
 This bias serves to calculate the thermal conductance of the {\it junction}. We assess the fluctuation symmetry in relation to this temperature bias, which in practice is very close to the imposed bias $\Delta T$. 
 Additionally, $\Delta T_\text{S-S}$ quantifies the temperature difference observed between the two sulfur atoms situated at the edges of the carbon chain. Unlike  $\Delta T_\text{Au-Au}$, which pertains to the metal-molecule-metal junction, $\Delta T_\text{S-S}$
 is used to calculate the thermal conductance of the molecule itself. 
We illustrate and discuss these temperature biases below in Fig. \ref{Fig4}(a)). 

The cumulative heat exchange data collected is used to generate the full probability distribution of heat exchange, for testing the steady state heat exchange fluctuation symmetry (\ref{eq:PQ}) and calculating cumulants of the current \cite{Naim}.
Using simulation data, we generate the ensemble of heat exchange values during time interval  $\tau$. For example, for the hot side
$Q^{\text{hot},(i)}_{\tau} = \Delta E_{\text {hot}}(t_i + \tau) - \Delta E_{\text{hot}}(t_i)$. 
We generate three such sets based on heat exchanged at the cold  and hot baths, as well as when using the averaged heat exchanged  $\Delta E_\text{ave}(t)$, to yield  $Q_{\tau}^\text{cold}$,
$Q_{\tau}^\text{hot}$, and
$Q_{\tau}^\text{ave}$, respectively.
%
For these three ensembles 
we generate histograms of the heat exchange, $P(Q_{\tau})$, and test the fluctuation symmetry, Eq. (\ref{eq:PQ}). Such results are presented in Fig. \ref{Fig5} for the many-atom thermostats.
We turn to single-atom thermostats in Fig. \ref{Fig6}. We present the histogram of the heat exchange in Fig. \ref{Fig7} and we use the probability distribution of heat exchange to compute the current noise and the TUR in Fig. \ref{Fig8}. 
As for the values of $\Delta\beta$ in Eq. (\ref{eq:PQ}), rather than substituting the thermostats' target temperatures, $T_\text{hot}$ and $T_\text{cold}$, we use here the actual steady state temperatures developing at the metal right next to the molecule. 


\begin{figure}[htbp] 
    \centering
    \includegraphics[width=\columnwidth]{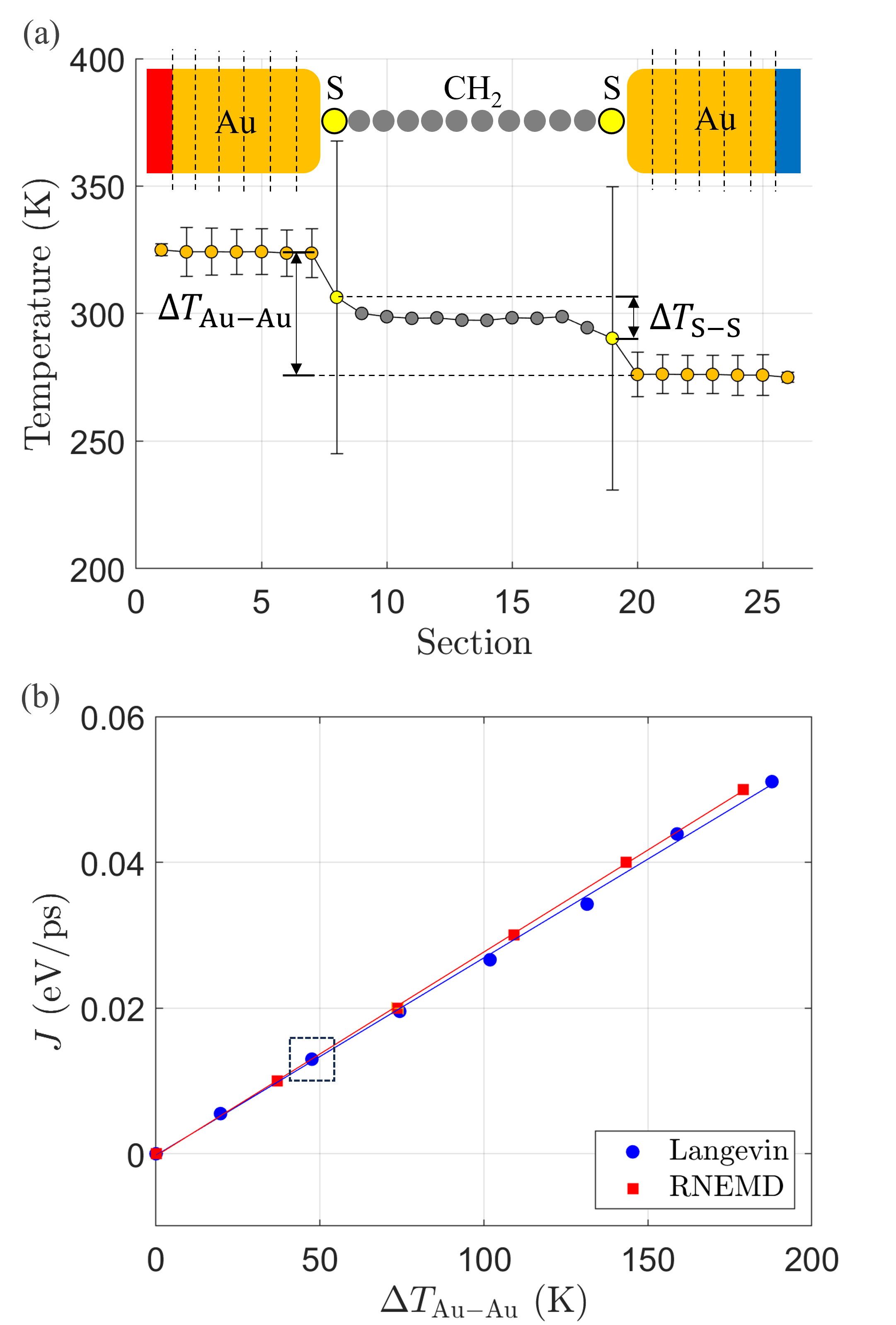}
    \caption{ {\bf Calculations of the heat current with different methods.}
    {\bf (a)} An example of temperature profile generated on the junction in Langevin NEMD simulations. The imposed temperature difference is $\Delta T=50$ K and we mark the generated biases, $\Delta T_\text{Au-Au}$ and $\Delta T_\text{S-S}$. The values of temperatures on Au and S atoms are shown along with their standard deviation.
    {\bf (b)} Heat current $J$ as a function of temperature difference of the junction $\Delta T_\text{Au-Au}$,
    comparing the Langevin NEMD (focus of this work) and RNEMD \cite{JW-HeatMD} simulations.  In both cases, we maintained $\bar{T} = 300$ K. The temperature profile for the data point highlighted by the dashed square appears in panel {\bf (a)}.
    For the Langevin NEMD simulations, a damping time $\gamma^{-1} = 0.04$ ps was applied. 
    Simulations were repeated three times each. The size of the resulting error bars appear less than the size of markers.}
    \label{Fig4}
\end{figure}

\section{Conventional many-atom thermostats}
\label{many-sec}

\subsection{Average heat current}
\label{G-Sec}

We begin with the calculation of the thermal conductance $G$ through NEMD simulations.
In a previous study \cite{JW-HeatMD}, we tested two methodologies for assessing thermal transport within junctions: The approach-to-equilibrium MD (AEMD) method, which extracts the thermal conductance from the equilibration dynamics \cite{AEMD},
and the reversed NEMD (RNEMD) method, where we impose the steady state current passing through the junction, and recover the internal temperature bias \cite{Pawel11,Luo10,Shub15,Shub17}.  
In Ref. \citenum{JW-HeatMD} we demonstrated that both methods provided results consistent with other studies and with experiments.
The RNEMD method in particular came up as a robust tool. However, since the RNEMD method pins the heat current, it cannot be used to obtain the current fluctuations.

In the present work, however, we are interested in the statistics of heat exchange. As such, we have chosen to adopt a "direct" NEMD technique. In this method, the Langevin thermostats dictate the temperatures at the boundaries, and the heat current develops in accord.
However, before proceeding to explore the heat exchange statistics, the method necessitates additional testings and verification. Two specific aspects of the method 
warrant examinations: 
(i) Determining the region of gold that should be subjected to thermalization. Intuitively, one would like to ensure that the region that is coupled to the bath is sufficiently distant from the junction of interest, thus avoiding interference of the Langevin thermostat with the actual phenomena under investigation.
(ii) Optimizing the Langevin damping parameter $\gamma$. This thermostating parameter dictates the rate at which local thermal equilibrium is achieved. Striking the right balance is important: Small $\gamma$ values lead to slow thermalization but minimize interference of the thermostat with the intrinsic dynamics. Large $\gamma$ values accelerate thermalization, but at the potential cost of modifying the intrinsic system's behavior.

We examine both of these aspects in Fig. \ref{Fig3}.
We simulate two scenarios, as illustrated in Fig. \ref{Fig3}(a): The Langevin baths are either coupled to the termini of the mobile Au segment (top) or they are connected to the entire moving Au atoms (bottom), as was done in e.g., Ref. \citenum{Nitzan22}.
In both cases, we conduct simulations of the heat current, and we determine the thermal conductance of the junction by considering the temperatures 
taken from the slab of gold atoms closest to the junction. 

Fig. \ref{Fig3}(b) indicates that there is a significant difference between these two scenarios,
depending on the value of the damping parameter.
Note that the damping parameter, corresponding to the inverse of $\gamma$, is given here in units of time. A smaller value of the damping parameter signifies frequent interactions with the Langevin bath to regulate temperature, and vice versa \cite{LAMMPS}. 
When the damping time is short, having the entire Au system coupled to baths (hot or cold) results in conductance that is dependent on $\gamma$  (empty circles). This however is not desirable; the implementation of the thermostats should not affect the conduction of the junction.
In contrast, we observe that the conductance remains independent of the application of the Langevin baths when they are applied only to the terminal regions of the gold (full circles). 

Our calculation of the Au-alkanedithiol-Au system consistently results in thermal conductance of the junction of approximately 40-45 pW/K, in agreement with previous studies \cite{JW-HeatMD}. Notably, experiments have reported  a smaller value in the range of 20 pW/K \cite{CuiExp19}. 

Altogether, Fig. \ref{Fig3} shows that care needs to be exercised when employing Langevin baths in NEMD simulations for thermal transport. Moving forward, we have chosen to adopt the configuration with baths applied to the ends of the Au system (setup (i) in Fig. \ref{Fig3}(a)). Additionally, we have selected a damping time of 0.04 ps as our optimal parameter for ongoing simulations.

To further examine our calculations, we simulate the heat current at various temperature biases 
using both the Langevin-based NEMD and the RNEMD method as described in Ref. \citenum{JW-HeatMD}.
First, in Fig. \ref{Fig4}(a) we illustrate the temperature profile generated in the Langevin-based NEMD method at $\Delta T=50$ K.
The temperature  drops at the boundaries, while it is about constant in this quasi one-dimensional molecule. This indicates that the interfacial (molecule-metal) thermal resistance is the main obstacle to conduction in this junction \cite{Baowen22}, while the molecule conducts approximately ballistically. 
Next, in Fig. \ref{Fig4}(b) we show that both methods exhibit a linear relationship between the heat current and the temperature difference in the junction. 
This trend is expected for alkane chains, and the two methods follow each other well. 
Note that $\Delta T_\text{Au-Au}$ represents the steady state temperature difference on gold segments next to the molecule,
differing marginally from the thermostat temperature difference, $\Delta T$, as can be seen in Fig. \ref{Fig4}(a). 
As a reminder, in the RNEMD method, the steady state temperature difference is deduced from the temperature profile that develops on the junction. In the present "direct"  Langevin NEMD method, $\Delta T_\text{Au-Au} \leq \Delta T$; a small temperature gradient develops across the Au leads due to thermostats placed at the ends. 
The thermal conductance $G$ can be extracted easily from the current-bias slope to be 43.4 and 44.8 pW/K for the NEMD and RNEMD methodologies, respectively.

\begin{figure*}[tb] %
    \centering
    \includegraphics[width=2\columnwidth]{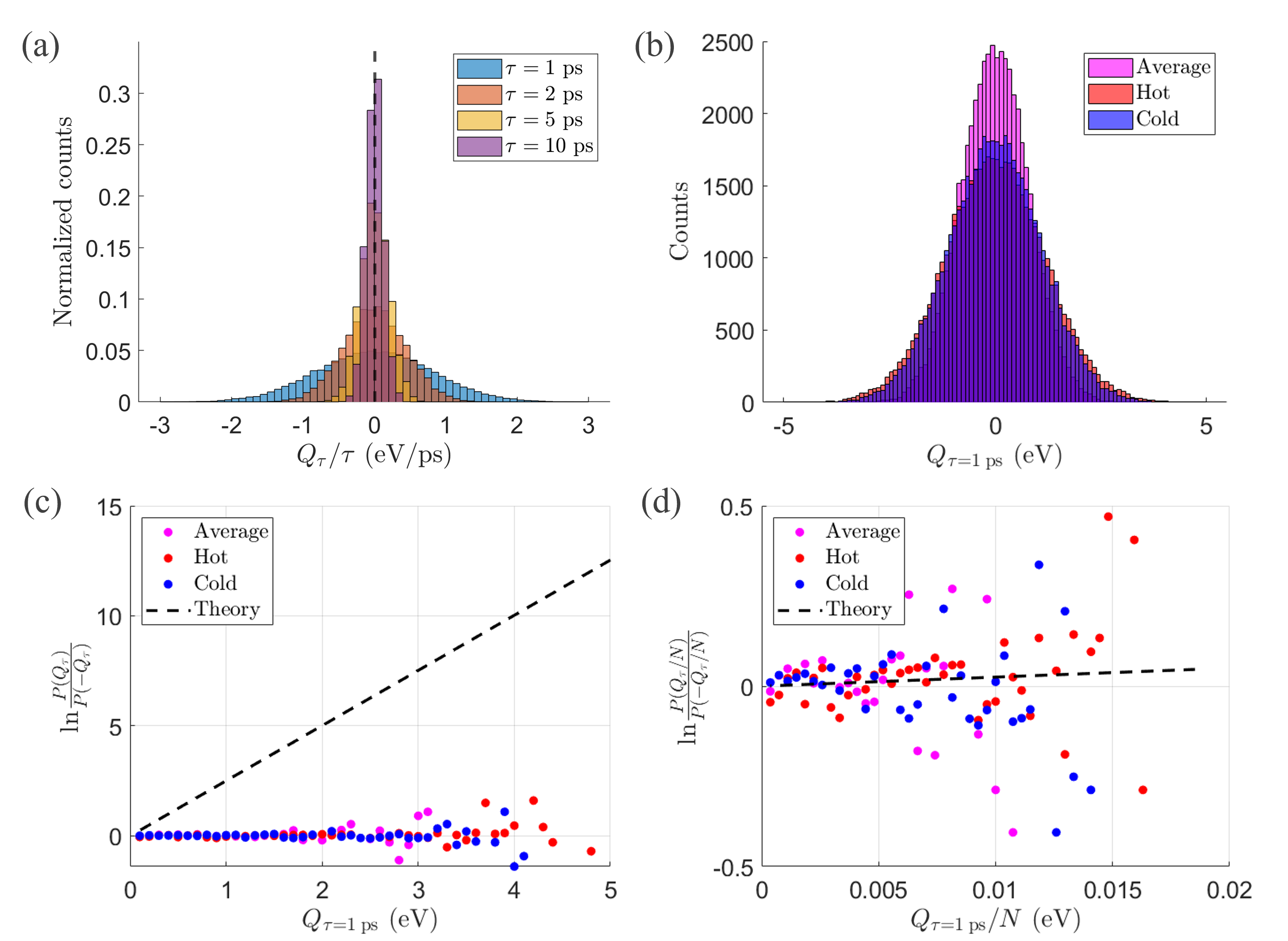}
    \caption{{\bf Testing the fluctuation symmetry in many-atom thermostats. 
    (a)} Histograms of integrated heat currents $Q^\text{ave}_{\tau}/\tau$ with varying time interval $\tau$. The black dotted line marks the value of the mean heat current $J$. 
    {\bf(b)} Overlaid histograms of integrated heat currents from individual baths, $Q^\text{hot}_{\tau}$ and $Q^\text{cold}_{\tau}$, and $Q^\text{ave}_{\tau}$ with $\tau = 1$ ps. 
    The histograms contain 50,000 data points for 50 ns simulation time, with $\tau = 1$ ps being the lowest time interval between each data points.
     {\bf(c), (d)} Fluctuation symmetry plot with (c) $Q_{\tau}$ and (d) $Q_{\tau}/N$, where $N = 270$ Au atoms coupled to thermostat on one end. Dashed lines indicate the theoretical $\Delta\beta = 2.5$ eV$^{-1}$. Simulation parameters are imposed temperatures  $T_\text{hot}=310$ K and $T_\text{cold}=290$ K, with
     $\Delta T_\text{Au-Au}\approx 20$ K, 
     $\gamma^{-1} = 0.04$ ps.
     }
    \label{Fig5}
\end{figure*}

\subsection{Fluctuations}
\label{Fluc-sec}

We explore fluctuations in heat transport by examining the ensemble of $Q_{\tau}$, the net heat exchange between the baths and the system, occurring within the time interval $\tau$. We then construct the probability 
distribution function by generating histograms, as explained in Sec. \ref{Method-sec}.
In simulations, we test different values for $\tau$, from 1 to 10 picoseconds, and separately study heat exchange at either the hot or cold contacts, as well as the behavior of the averaged heat exchanged. 

Fig. \ref{Fig5} summarizes the analysis of the fluctuation data. 
In Fig. \ref{Fig5}(a) we illustrate how the probability distribution function evolves as a function of the inspected time interval $\tau$. 
Note that upon increasing $\tau$, we have fewer data points available for the analysis.
Plotting the histogram against $Q_{\tau}/\tau$,  the distributions become narrower as  $\tau$ increases, but the positions of the average remain fixed.
We extract from Fig. \ref{Fig5}(a) the mean heat current, which is approximately 0.005 eV/ps, or  $\approx 800$ pW. This value gives the expected thermal conductance $G = 40$ pW/K with $\Delta T_\text{Au-Au} \approx 20$ K.

In Fig. \ref{Fig5}(b), we compare the probability distribution functions calculated at the hot and cold ends, denoted by $Q^\text{hot}_{\tau}$ and  $Q^\text{cold}_{\tau}$, respectively, along with 
a histogram generated by averaging the heat exchange process, $Q^\text{ave}_{\tau}$. 
We find that the  distribution generated from $Q^\text{hot}_{\tau}$ is slightly wider 
than that created from $Q^\text{cold}_{\tau}$.  However, the mean heat currents for these different histograms are identical. To test the heat exchange fluctuation symmetry, Eq. (\ref{eq:PQ}), we investigate the ratio $\ln [P(Q_{\tau})/P(-Q_{\tau})]$ for each of the three sets, presented in Fig. \ref{Fig5}(c).
Remarkably, results dramatically deviate from the expected slope of $\Delta \beta$.
Instead, each histogram behaves as almost an equilibrium distribution, and it does not reflect the nonequilibrium heat exchange process taking place.
%
We attribute this failure of the NEMD simulations to respect the fluctuation symmetry to several issues: 

 {\it 1. Choice of the integrator}: The velocity Verlet integrator 
 employed by LAMMPS  to solve the Langevin equation is not suitable for solving stochastic differential equations. Rather, it treats the random force on an equal footing as deterministic atomic forces. 
 This in turn leads to substantial errors when
using the interface definition for heat current  Eq. (\ref{eq:Jdef}) as discussed in  Ref. \citenum{Naim}. 
Other studies noted related problems in the velocity-Verlet or other common MD integrators resulting in difficulties generating 
correct equilibrium properties.
Revisions to the standard velocity-Verlet scheme were implemented in LAMMPS, by coupling the deterministic and stochastic forces \cite{GJF1}. These modified schemes generate the correct Boltzmann distribution \cite{GJF1,GJF2,GJF4} and satisfy the Green-Kubo relation \cite{GJF3}.
However, we tested the scheme of Ref. \citenum{GJF1} and found that it did not resolve the fluctuation symmetry issue we observed with many-atom thermostats, and it produced results that were similar in trends to Fig. \ref{Fig5}(c).

In addition, LAMMPS provides a method for calculating heat flux through the molecule (in a vector form) by computing the energy and stress tensor  
between interacting atoms  \cite{LAMMPS}. 
This method still did not produce the correct fluctuation symmetry while introducing further ambiguities, such as in the  estimation of the molecular length and in choosing the directionality of heat transport,
resulting in an incorrect estimate for the average current. 

{\it 2. Definition of heat current:} 
As we demonstrated in Ref. \citenum{Naim} on a one-dimensional model system, the "interface" heat current definition, Eq. (\ref{eq:Jdef}),  is difficult to converge to the correct result, unlike an
"internal" definition that counts the amount of energy exchanged from internal, interatomic interactions. Moreover, we showed in Ref. \citenum{Naim} that while the internal definition for heat exchange obeyed the fluctuation symmetry, even with a rough time step, the definition (\ref{eq:Jdef}) in practice dissatisfied this symmetry, and even when the averaged currents converged to the correct value.


We emphasize that mathematically, there is nothing wrong with Eq. (\ref{eq:Jdef}). Rather, as we showed in Ref. \citenum{Naim} this expression is slower to converge to the long time limit than the internal definition as one reduces the time step. The interface definition is also extremely sensitive to the integrator used and how exactly the stochastic force is evaluated, showing violations to the fluctuation symmetry even at very small timesteps \cite{Naim}. 

 
{\it 3. Many-atom thermostats:} We further suspect that
the application of many thermostats at each side, contributes to the generation of an effective local equilibrium at those ends, rather than a true nonequilibrium steady state.
To probe this aspect,
 in Fig. \ref{Fig5}(d) we test the fluctuation symmetry by scaling the heat, $\tilde Q_{\tau}=Q_{\tau}/N$, 
 with $N= 270$ the number of thermostated atoms at each edge. Remarkably, the relationship $P(\tilde Q_{\tau})/P(-\tilde Q_{\tau}) = e^{\Delta \beta \tilde Q_{\tau}}$ appears to hold for the scaled current. This rough observation
 motivates us to explore the transport behavior using {\it single-atom thermostats} as means to generate the correct statistics of heat exchange. This idea is developed and detailed in the next section.






\begin{figure*}[htbp]
    \centering
    \includegraphics[width=2\columnwidth]{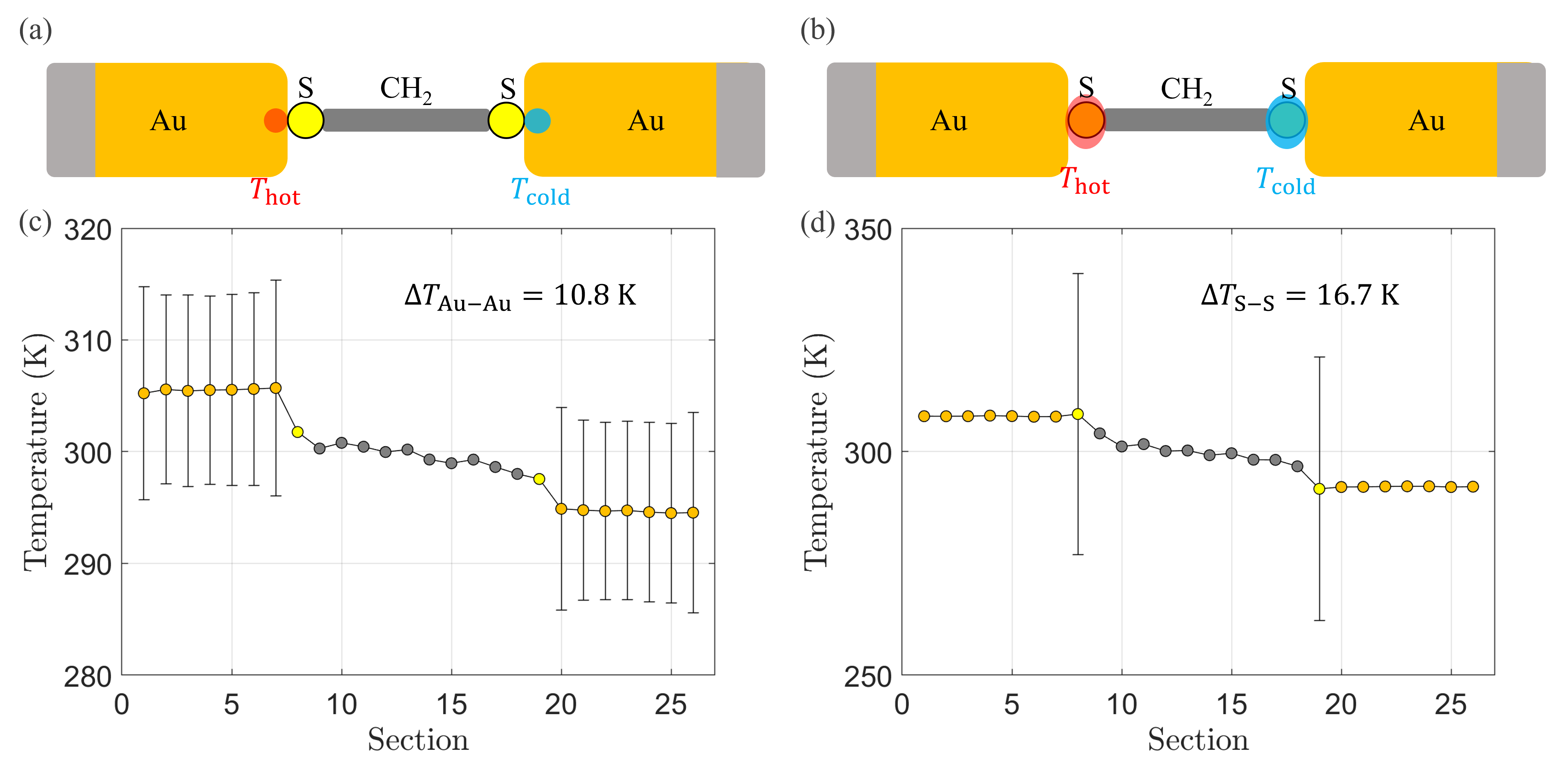}
    \caption{{\bf Single-atom thermostats coupled to single atoms on both leads and the resulting temperature profile on the junction.   }
    Illustration of thermostat positions are shown for {\bf(a)} thermostats coupled to single Au atoms at the interface between lead and molecule, {\bf(b)} thermostats coupled to single S atoms. {\bf (c), (d)} Temperature profile plots of thermostat setups under respective illustrations. For both systems, simulation parameters are $\bar{T} = 300$ K, $\Delta T = 20$ K, $\gamma^{-1} = 0.04$ ps.
    }
    \label{Fig6}
\end{figure*}


\begin{figure*}[htbp]
    \centering
    \includegraphics[width=2\columnwidth]{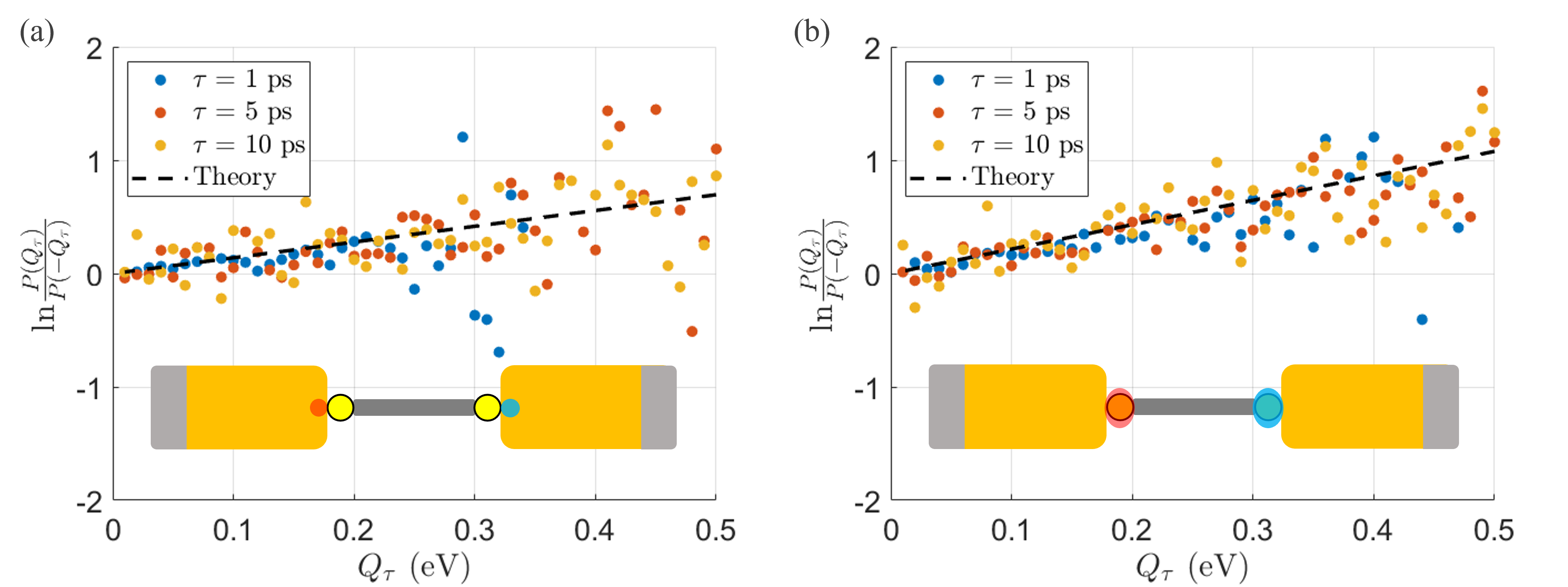}
\caption{{\bf Fluctuation symmetry analysis of Langevin thermostats coupled to single atoms on both leads.} {\bf (a)} Fluctuation symmetry plot with a single gold atom thermalized at each side. The dashed line depicts  the theoretical prediction with the slope $\Delta\beta = 1.39$ eV$^{-1}$. 
{\bf (b)} Same, but thermalizing the S atoms at the ends of the alkane chain. The dashed line depicts the theoretical linear prediction with the slope $\Delta\beta = 2.16$ eV$^{-1}$. 
For both systems, the distributions contain 50,000 $\tau=1$ ps data points and each element is constructed as
$Q^{\text{ave},(i)}_{\tau}= [Q^{\text{cold},(i)}_{\tau} +Q^{\text{hot},(i)}_{\tau}]/2$.
Simulation parameters are $\bar{T} = 300$ K, $\Delta T = 20$ K, $\gamma^{-1} = 0.04$ ps.
}
    \label{Fig7}
\end{figure*}

\section{Single-atom thermostats}
\label{single-sec}

\subsection{Fluctuation symmetry}

In an effort to generate data consistent with the fluctuation symmetry, we conduct simulations in which we implement single-atom thermostats. In this configuration, only a single atom on each side of the junction is coupled to the (hot or cold) thermostats. A graphical representation of this system is shown in Fig. \ref{Fig6} (a)-(b). 

First, in Fig. \ref{Fig6}(a), we illustrate the setup in which we thermalize a single gold atom at each side of the system, one towards a high temperature (310 K) and the other to a lower temperature (290 K). The temperature profile developing in the junction is displayed in Fig. \ref{Fig6}(c). We note that the actual temperature bias is lower than 20 K, which is reasonable given the "gentle" nature of single-atom thermostats in this nonequilibrium (two-bath) setup. Calculating the actual temperature bias as developing on the gold section attached to the molecule we get $\Delta T_\text{Au-Au}=10.8$ K. 

The corresponding fluctuation symmetry (\ref{eq:PQ}) is examined in Fig. \ref{Fig7}(a), and it is verified when compared to the theoretical curve with the generated $\Delta T_\text{Au-Au}$. 
In the Appendix, we study the heat exchange symmetry when evaluated at each contact.
Notably, only once the heat is averaged over the two sides, as we do in Fig. \ref{Fig7}(a), a single-atom thermostat provides data consistent with the fluctuation symmetry.  

We also look at the case in which the thermostats are applied onto the sulfur atoms that contact the alkane chain to the gold, as shown in Fig. \ref{Fig6}(b). The resulting temperature profile is displayed in Fig. \ref{Fig6}(d) with $\Delta T_\text{S-S}=16.7$ K. 
The fluctuation symmetry analysis for this configuration is presented in the Appendix for the individual baths, and  in Fig. \ref{Fig7}(b), once averaging heat exchange over the two baths.
Note that when applying the thermostats directly onto the S atoms, we extract the molecular conductance characterizing the alkanedithiol chain, rather than the overall junction's conductance.
 The heat current is evaluated to be 0.0125 eV/ps. The calculation of the molecular conductance $G_\text{M}$ using $\Delta T_\text{S-S}$ yields approximately 120 pW/K, which agrees with previous results \cite{JW-HeatMD}. 

As for the fluctuation symmetry, in both applications of single-atom thermostats, either on a gold atom or on the sulfur, we find that it is obeyed when using the set $Q^\text{ave}_{\tau}$. 
 We further test the fluctuation symmetry for different time intervals, long enough to wash out initial conditions. We confirm in Fig. \ref{Fig7} that the fluctuation symmetry is valid irrespective of the chosen (long) time interval. Note however that we have a fixed-size set of 50,000  $\tau=1$ ps data points. Using
longer time intervals reduce the ensemble size, e.g., 
  we have only 5,000 $\tau=10$ ps data points. 

It is important to emphasize that our construction of single-atom thermostats is not meant to suggest that this is how experiments are to be conducted. Rather, single-atom thermostats serve only as a {\it  computational} device to address the shortcoming of the conventional NEMD simulation method. Its deficiencies stem from the improper solver and the heat current definition, exacerbated by having many atoms thermalized in conventional Langevin NEMD calculations.

In sum, conventional NEMD simulations as implemented in LAMMPS fail to reproduce data consistent with the heat exchange fluctuation symmetry.
As an ad-hoc solution, we produced the correct statistics by (i) applying the cold and hot thermostats each on a single atom. (ii) Generating data after averaging the heat exchange at the two sides.
A fundamental cure to this deficiency would be to use stochastic integrators and adopt a different definition of heat exchange: Rather than calculate it as a net heat exchange with the bath, evaluate the heat current across bonds, internally \cite{Naim}.




\begin{table*}[htbp]
    \begin{center}
    \caption{Table of thermal conductance values as extracted from different thermostat configurations}
	\footnotesize
	\begin{tabular}{|m{5em}|m{7em}|m{5em}|m{5em}|m{6em}|m{5em}|m{5em}|}
		\hline
		 \bf    Imposed target $\Delta T$ (K) & \bf Thermostat setup  & \bf Measured $\Delta T_\text{Au-Au} $ (K) & \bf Measured $\Delta T_\text{S-S}$ (K) & \bf Averaged heat current $J$ ($\times 10^{-3}$ eV/ps) & \bf Junction thermal conductance $G$ (pW/K) & \bf Molecular thermal conductance $G_\text{M}$ (pW/K)\\ 
		\hline
		 \multirow{3}{*}{20} & Au atoms at ends  & 19.7  & 8.50  & 5.50  & 44.8  & 104 \\
		\cline{2-7}
& Single Au atom  & 10.8  & 4.21  & 3.40 & 50.6  & 130 \\
                \cline{2-7}
& Single S atom  & 15.8 & 16.7 & 12.5 & - & 120\\
		\hline
\hline
		 \multirow{3}{*}{50} & Au atoms at ends  & 47.5  & 16.1  & 13.0  & 43.7  & 129 \\
		\cline{2-7}
& Single Au atom  & 28.2  & 10.5  & 7.50 & 42.3  & 114 \\
                \cline{2-7}
& Single S atom  & 40.2 & 45.2 & 31.4 & - & 111\\
		\hline
  \hline
		 \multirow{3}{*}{80} & Au atoms at ends  & 74.3  & 25.1  & 19.5  & 42.1  & 125 \\
		\cline{2-7}
& Single Au atom  & 45.0  & 15.6  & 12.2 & 43.5  & 126 \\
                \cline{2-7}
& Single S atom  & 61.6 & 64.1 & 50.0 & - & 125\\
		\hline
	\end{tabular}
\label{Table1}
\end{center}
\end{table*}








\begin{figure*}[htbp]
\centering
    \includegraphics[width=2\columnwidth]{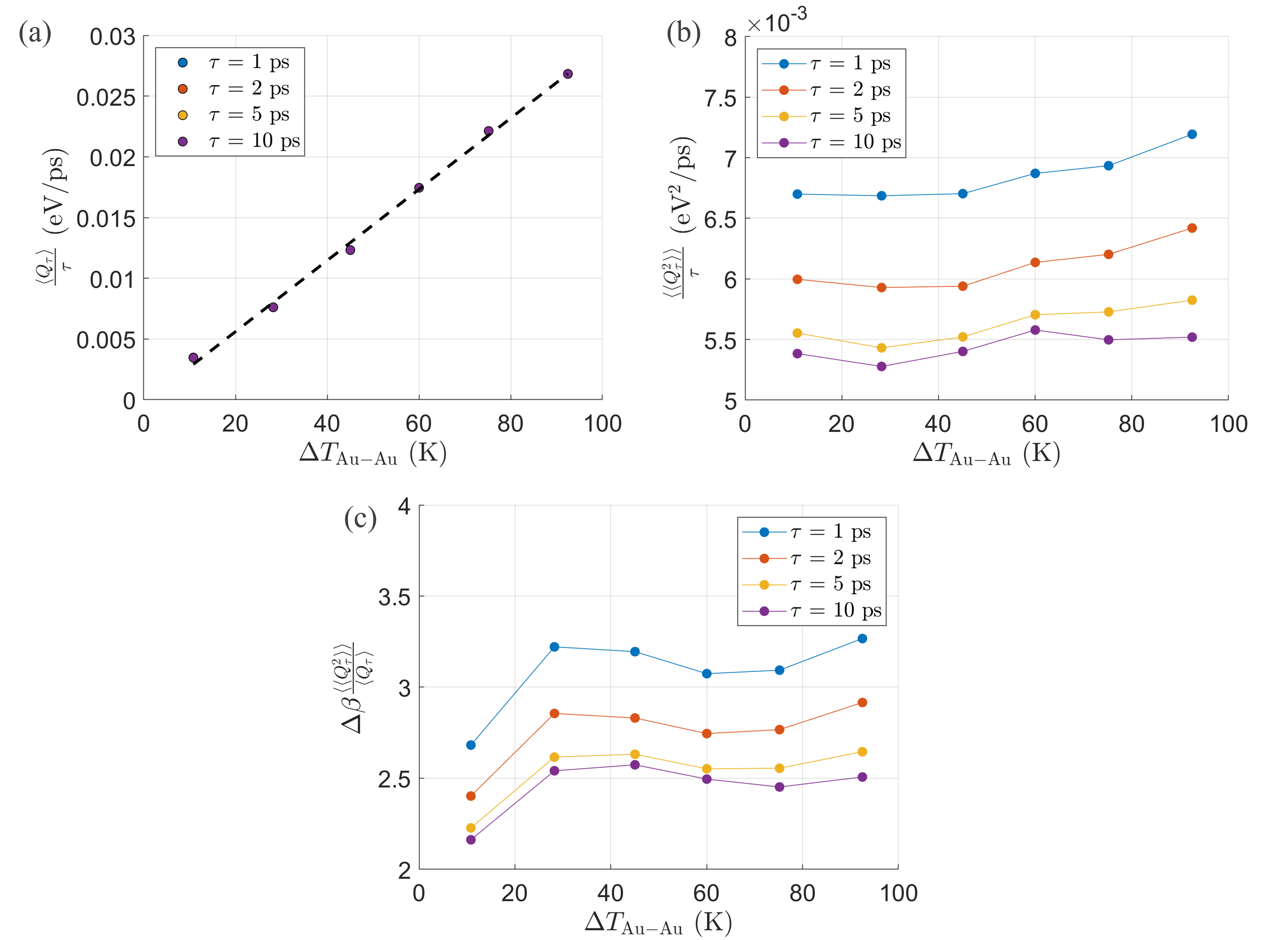}
\caption{ {\bf Heat current and its noise with single-atom thermostats.} Data is 
plotted against the 
temperature difference developing between Au interface at the boundary of the molecule, $\Delta T_\text{Au-Au}$.
{\bf (a)} Averaged heat current, $\langle Q_{\tau}\rangle/\tau$ (data overlapping), 
{\bf (b)} heat current variance, $\langle\langle Q_{\tau}^2\rangle\rangle/\tau$
and {\bf (c)} the TUR ratio, for which theoretical TUR ratio is greater or equal to 2.
Results were generated from the ensemble $Q_\tau^\text{ave,(i)}$ and are presented when using several different time intervals, $\tau$.
}
 \label{Fig8}
\end{figure*}
\subsection{Conductance and the TUR}


In the previous subsection, we demonstrated that a single-atom thermostat allowed generating data that respected the fluctuation symmetry. Building upon this result, we employ such thermostats to compute the thermal conductance of both the junction and the molecular building block, and confirm that they act properly.
Furthermore, we study the behavior of the heat current and current noise as a function of temperature bias.
Using the current cumulants, we also investigate the thermodynamic uncertainty relation \cite{TURUdo1}. 

In the Table \ref{Table1} we present calculations of the thermal conductance obtained from three approaches:
(i) Many-atom thermostats, acting on remote segments of the gold (270 atoms) as illustrated in Fig. \ref{Fig1}.  
(ii) Single-atom thermostat applied onto a single gold atom placed next to the molecule.
(iii) Single-atom thermostat, acting on the S atoms thus directly on the molecule. Note that this latter approach does not allow for the calculation of the junction's thermal conductance, $G$. 

We find that the different methods yield consistent results for the thermal conductance, particularly when the imposed temperature bias is large. In contrast, when the imposed bias is small, the resulting $\Delta T_{\text{Au-Au}}$ is even smaller, and relative errors in the calculation increase.
Overall, we conclude that single-atom thermostats are a suitable tool for calculating the thermal conductance of molecular junctions. 

In Fig. \ref{Fig8}, we study the behavior of the heat current, its noise, and the TUR as a function of temperature bias. These results were obtained by studying the set $Q_\tau^\text{ave}$. 
First, in Fig. \ref{Fig8}(a) we present the average current, computed by averaging over the ensemble of $Q_{\tau}$ and dividing it
 by the time interval considered, which varies from 1 ps to 10 ps. We find that the current remains the same across these different choices of time intervals, manifesting that the time interval is sufficiently long to achieve steady state. 
 
Next, we turn in Fig. \ref{Fig8}(b) to the second cumulant of $P(Q_{\tau})$, denoted as 
$\langle \langle Q_{\tau}^2 \rangle \rangle  = \left[ \langle Q_{\tau}^2\rangle - \langle Q_{\tau}\rangle^2  \right]$.  
The current noise, or the scaled fluctuations of heat exchange  $\langle \langle Q_{\tau}^2 \rangle \rangle/\tau $ is shown to approach a constant value with $\tau$ when $\tau$ is on the order of 10 ps.

Finally, we probe  in Fig. \ref{Fig8}(c) the TUR inequality,
\bea
 \frac{\langle \langle Q_{\tau}^2  \rangle \rangle}{\langle Q_{\tau }\rangle^2} \frac{\langle \sigma_{\tau}\rangle}{k_B} \geq 2,
 \eea
with $\langle \sigma_{\tau} \rangle$ as the cumulative entropy production during the time interval $\tau$, which is long enough for steady state to set in.
This dissipation-precision inequality translates to 
 \bea
\Delta \beta 
 \frac{\langle  \langle Q_{\tau}^2 \rangle \rangle}{\langle Q_{\tau }\rangle}  \geq 2
 \label{eq:TURh}
 \eea
for heat transport problems as we study here.
The relationship (\ref{eq:TURh}) has been proved for certain classes of models and dynamics: Continuous time Markov state process \cite{TURGing}, overdamped Langevin dynamics \cite{TURLang,DechantSasa}, and fully harmonic models \cite{TURheat}. 
It was also examined experimentally in Ref. \citenum{TURhE}, displaying violations to Eq. (\ref{eq:TURh}) due to quantum non-Markovian effects.
While previous studies examined the TUR in simplified model systems, this relationship has 
rarely been assessed with realistic atomistic simulations \cite{Roldan}. Specifically, it has not been studied in a physical molecular junction setup including realistic-anharmonic interactions, which we do here. 
Moreover, since Langevin thermostats act here on a single atom, the dynamics maintains inertial effects, thus it cannot be captured by overdamped equations. 

In Fig. \ref{Fig8}(c), we display the TUR ratio, corresponding to the left-hand side of Eq. (\ref{eq:TURh}), for several values of $\tau$. 
Close to equilibrium, the TUR ratio appears to approach the expected value of 2. However, with an 
increasing temperature bias, we observe that the TUR ratio first increases, but then saturates, 
yet obeying the inequality in Eq. (\ref{eq:TURh}).  
Despite the fact that the heat current displays a linear dependence on the temperature bias even in the far-from-equilibrium regime, the second cumulant may gain nontrivial dependence on $\Delta T_\text{Au-Au}$ as it grows. As such, the overall behavior is not necessarily characterized by linear response, accounting for the TUR's departure from its expected value of 2.
Note that using the Green-Kubo relation, $S_\text{eq}=\frac{J}{\Delta T} 2 k_B\bar T^2 $, with $S_\text{eq}$ the equilibrium noise, we find
that for thermal conductance around 44 pW/K, the equilibrium noise should be 4.2 $\times 10^{-3}$ eV$^2$/ps. Fig. \ref{Fig8}(b) shows higher values, indicating additional contributions due to the nonequilibrium situation, as is further evidenced by the TUR value exceeding two. 

Altogether, our simulations indicate that the TUR tradeoff is satisfied in this real-life model for a molecular junction within the examined temperature range, which is relevant for experiments.

%

\subsection{Experimental relevance}

The time interval considered in Fig. \ref{Fig7}(a) was in the 1-10 ps range with a temperature difference of about 10 degrees,
$\Delta \beta\approx 1.4$ eV$^{-1}$. This translates to currents in the 0.003 eV/ps range (see Table \ref{Table1}).
Considering a heat exchange event with $Q_{\tau} = 0.2$ eV, we find that  $P(Q_{\tau})/P(-Q_{\tau}) \approx 1.3$. Thus, we expect that one should frequently-enough observe negative entropy production at this amount.
However,  time resolution is limited experimentally. Let us now assume that heat exchange is probed in time intervals of nanoseconds. Recall that approximately, the heat current in our system
 is 3 eV/ns for $\Delta T_\text{Au-Au}=10$ K (see  Fig. \ref{Fig7}(a)). Within a ns time interval, a typical measurement would be in the $Q_{\tau}$ = 10 eV range (see Fig. \ref{Fig2}), translating into $P(Q_{\tau} )/P(-Q_{\tau} ) = 4 \times 10^5$. Thus, if the time resolution in measurements is in nanoseconds, tens of millions of experiments would be required in our system in order to observe an instance of heat exchange against the temperature bias. Even more so, for a fluctuation at the amount of 20 eV, the ratio of probabilities exceeds $10^{11}$.
Note that increasing the averaged temperature, e.g., using $T_\text{hot}=610$ and $T_\text{cold}=590$ K, allows stronger fluctuations that facilitate data collection towards analyzing the fluctuation symmetry.

With measurements of heat exchange and its statistics in single-molecule junctions being extremely challenging, the role of
simulations as we undertake in this study becomes even more important, serving to guide such efforts. As a first step, our work here focused on the assessment and development of fitting computational tools.

\section{Summary}
\label{Sum-sec}

In this work, we studied the statistics of heat exchange using an atomistic description for a nanojunction. We noted the challenge of satisfying the fluctuation symmetry within conventional computational methods, and suggested a solution to produce physical data for the heat current noise.

We utilized the conventional nonequilibrium molecular dynamics approach as implemented in LAMMPS, where segments of the metal contact are thermally controlled by Langevin baths. While this standard approach yields the correct behavior for the averaged current and the thermal conductance, we showed that it failed to generate the correct statistics of heat exchange. 

Our solution involved the implementation of single atom thermostats, acting on single atoms at each boundary, and the additional averaging of heat exchange from the hot and cold edges. This ad-hoc approach successfully generated data that was consistent with the fluctuation symmetry. 
Additionally, these calculations of thermal conductance were consistent with other established methods. From the calculation of the second cumulant of heat exchange we were able to examine the thermodynamic uncertainty relation, which we showed to hold in the alkane molecular junction.


It was proved in Ref. \citenum{TURheat} that the thermodynamic uncertainty relation (\ref{eq:TURh})
universally holds for harmonic systems in the steady state limit. 
In our study, the force field includes anharmonic interactions. For example, the S-Au bond is treated as a Morse potential, nonbonding interactions are handled with the Lennard-Jones potential, and dihedral terms are included within the molecule. 
However, at room temperature, heat transport in Au-alkane-Au chains is dominated by harmonic interactions resulting in ballistic transport \cite{Dvira2003,Lu2021}. 
As such, the observation that the TUR is obeyed in our system, as displayed in Fig. \ref{Fig8}, is not surprising.
It is intriguing to probe the TUR in highly anharmonic junctions, in the underdamped dynamical limit, to assess its applicability in anhamrmonic systems.

Classical MD simulations are an effective tool for capturing the interatomic anharmonic force field. However, quantum effects become important when vibrational molecular modes and substrate phonons are of a higher frequency than the thermal energy.
To treat, e.g., molecular junctions attached to Graphene electrodes, semiclassical methods 
were developed, incorporating quantum statistics into the modified Langevin equation \cite{Wang07,Lu2021,LuGLE,Nitzan23}. 
Extending our work, it is important to test whether such hybrid simulation approaches respect the universal entropy production fluctuation symmetry both analytically and numerically. 



In future work we plan to investigate the thermal conductance and noise properties of more complex molecules, employing ab-initio MD or machine learning force fields rather than empirical potentials. 
However, clearly, 
future studies of current noise with molecular simulators
require implementing robust and rigorous stochastic solvers and heat current calculations.
 Once implemented, noise calculations will assist the design of robust and optimal thermal devices such as thermal conductors, insulators, diodes, and refrigerators.

\begin{acknowledgements}
DS acknowledges support from an NSERC Discovery Grant and the Canada Research Chairs program.
The work of MG was supported by an NSERC Canada Graduate Scholarship---Doctoral (CGSD). 
We acknowledge Longji Cui for discussions that motivated this work.
\end{acknowledgements}

\vspace{5mm}
\noindent {\bf AUTHOR DECLARATIONS} \\

\noindent {\bf Conflict of Interest}

\noindent The authors have no conflicts to disclose

\vspace{5mm}

\noindent {\bf DATA AVAILABILITY} \\
The data that support the findings of this study are available from the corresponding author upon reasonable request.

\begin{figure*}[tbp]
    \centering \includegraphics[width=2\columnwidth]{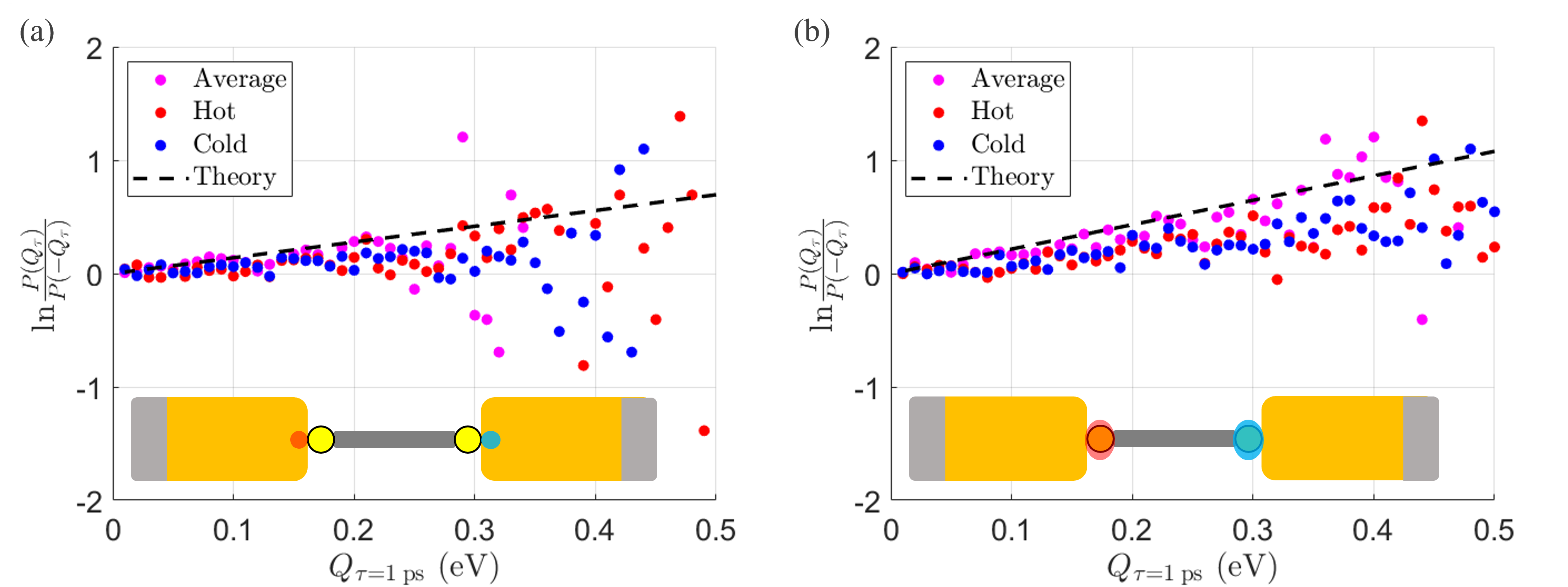}
    \caption{
    Fluctuation symmetry plots shown for {\bf (a)} thermostats coupled to single Au atoms and {\bf (b)} thermostats coupled to S atoms; the thermostats setups follow Fig. \ref{Fig6}. The analysis here is completed for individual baths, $Q^\text{hot}_{\tau}$ and $Q^\text{cold}_{\tau}$, in addition to $Q^\text{ave}_{\tau}$ using $\tau=1$ ps. The results are processed from the same distributions as in Fig. \ref{Fig6}-\ref{Fig7}, with simulation parameters $\bar{T} = 300$ K, $\Delta T = 20$ K, $\gamma^{-1} = 0.04$ ps.
    }
    \label{Fig9}
\end{figure*}
\renewcommand{\theequation}{A\arabic{equation}}
\renewcommand{\thesection}{A\arabic{section}}
\renewcommand{\thesubsection}{A\arabic{subsection}}
\setcounter{equation}{0}
\setcounter{section}{0} 
\setcounter{subsection}{0} 
\section*{Appendix: Fluctuation symmetry for single-atom thermostats }\label{app:1}

This Appendix discusses Fig. \ref{Fig9}, complementing Fig. \ref{Fig7}.
We use here the single-atom Au thermostat. We generate a set of $Q_{\tau}$,  build the full probability distribution function $P(Q_{\tau})$, 
and then test the fluctuation symmetry, separately for heat exchange data collected at the hot and cold baths. While these results should be identical, we do find in Fig. \ref{Fig9} that the statistics at the two baths is different.
However, we also build the histogram of the averaged heat exchanged at the two bath, 
$Q^{\text{ave},(i)}_{\tau} = [Q^{\text{cold},(i)}_{\tau} +Q^{\text{hot},(i)}_{\tau}  ]/2$. 
As we show in Fig. \ref{Fig7} in the main text, this combined data more closely follows the fluctuation ratio line. 

\end{document}